\documentclass{llncs}
\usepackage{epsfig}
\usepackage{amssymb,amsmath}
\begin{document}
\title{Persistence in voting behavior: stronghold dynamics in elections.}
%\title{Study of the strongholds in US Presidential Elections: confronting simulation with real data results.}
% \title{(1)\{Confronting real Electoral data results with Social Influence and Recurrent Mobility model.\} \\
% (2)\{Study of the strongholds in US Presidential Elections: confronting simulation with real data results.\}}
%\titlerunning{}  % abbreviated title (for running head)
\author{
Toni P\'erez \and Juan Fern\'andez-Gracia \and Jose J. Ramasco \and V\'ictor 
M. Egu\'iluz}
%\authorrunning{Toni P\'erez et al.} % abbreviated author list (for running head)
%\tocauthor{Toni P\'erez \and Juan Fernandez-Gracia \and Jose Javier Ramasco \and V\'ictor M. Egu\'iluz}
\institute{Instituto de F\'isica Interdisciplinar y Sistemas Complejos (IFISC)
\texttt{http://www.ifisc.uib-csic.es}\\
\email{toni@ifisc.uib-csic.es}
}

\maketitle 

\begin{abstract}
Influence among individuals is at the core of collective social phenomena such 
as the dissemination of ideas, beliefs or behaviors, social learning and the 
diffusion of innovations. Different mechanisms have been proposed to
implement inter-agent influence in social models from the voter model, to majority rules, to the Granoveter model.
Here we advance in this direction by confronting the recently introduced Social Influence and Recurrent Mobility (SIRM) model,
that reproduces generic features of vote-shares at different geographical levels,
with data in the US presidential elections.
Our approach  incorporates spatial and population diversity as inputs for the opinion dynamics 
while individuals' mobility provides a proxy for social context, and peer 
imitation accounts for social influence. The model captures the observed 
stationary background fluctuations in the vote-shares across counties. 
We study the so-called political strongholds, i.e., locations where the 
votes-shares for a party are systematically higher than average. A quantitative 
definition of a stronghold by means of persistence in time of fluctuations in 
the voting spatial distribution is introduced, and results from the US Presidential Elections 
during the period 1980-2012 are analyzed within this framework. We compare 
electoral results with simulations obtained with the SIRM model finding a good 
agreement both in terms of the number and the location of strongholds. The 
strongholds duration is also systematically characterized in the SIRM model. The 
results compare well with the electoral results data revealing an exponential decay in the 
persistence of the strongholds with time.

\end{abstract}

\section{Introduction}
People making a decision in a ballot are expected to follow a rational 
behavior. Rational arguments based on utility functions (payoff) have been 
considered in the literature regarding vote 
modeling~\cite{calculus_voting,estimating}. The rational hypothesis, however, 
tends to consider the individuals as isolated entities. This might actually be 
the reason why it fails to account for relatively high turnout rates in 
elections~\cite{Context,social_environment,social_calculus_voting}. The  
presence of a social context increases the incentive for a voter to actually 
vote, as he or she can influence several other individuals towards the same 
option~\cite{Fowler2005,Bond2012}. This effect is not only restricted to 
turnout, but also applies to the choices expressed in the 
election~\cite{zuckerman}. It is easy to find examples of people showing their 
electoral preferences in public in the hope of influencing their peers. Still 
social influence can also act in more subtle ways, without the explicit
intention of the involved agents to influence each other. The collective 
dynamics of social groups notably differs from the one observed from simply 
aggregating independent individuals~\cite{Lorenz2011}. Social influence is thus 
an important ingredient for modeling opinion dynamics, but it requires as well 
the inclusion of a social context for the individuals~\cite{Fowler2005,Bond2012,rogers_diff_innov,RevModPhys.81.591,Christakis_obesity,emotions_infectious,Rendell2010,Centola2010}.

Even though nowadays the pervasive presence of new information technologies has the 
potential to change the relation between distance and social contacts, we 
assume that daily mobility still determines social exchanges to a large extent. 
Human mobility has been studied in recent years with relatively indirect 
techniques such as tracking bank notes~\cite{Brockmann2006} or with more 
direct 
methods such as tracking cell phone communications~\cite{Gonzalez2008,Song2010}. A 
more classical source of information in this issue is the census. Among other 
data, respondents are requested by the census officers their place of residence 
and work. Census information is less detailed when considered at the individual 
level, but it has the advantage of covering a significant part of the 
population of full countries. Recent works analyzing mobile phone records have shown 
that people spend most of their time in a few locations~\cite{Song2010}. 
These locations are likely to be those registered in the
census and, indeed,  census-based information has been also used recently to 
forecast the propagation patterns of infectious diseases such as the latter 
influenza pandemic~\cite{Balcan2009,Balcan2009b,Sattenspiel95}.

In this work, we follow a similar approach and use the recurrent mobility 
information collected in the US census as a proxy for individual social 
context. 
This localized environment for each individual accounts mostly for face-to-face 
interactions and leaves aside other factors, global in nature, such as 
information coming from online media, radio and TV. Our results show that 
a model implementing face-to-face contacts through recurrent mobility and influence as 
imperfect random imitation is able to reproduce geographical and temporal patterns 
for the fluctuations in electoral results at different scales.

\section{Definitions and Methods}
We have investigated the voting patterns in the US on the county level. We have used the votes for presidential elections in years 1980-2012. 
For each county $i = 1, \dots, I$, we have data about the county geographic position, area, adjacency with other counties as well as data 
regarding population $N_{iy}$, and number of voters $V_{iyp}$ for each party $p$ for every election year $y=1,\dots,Y$ (note that $\sum_p V_{iyp} \neq N_{iy}$, 
since not everybody is entitled to vote). Raw vote counts are not very useful for comparing the counties as populations are distributed heterogeneously.
Therefore we switch from vote counts to vote shares
\begin{equation} v_{iyp} = V_{iyp} / \sum_p V_{iyp}~. \label{voteshare} \end{equation}

Since the votes received by parties others than Republicans and Democrats are minority, we have focused on the two main parties. 
We have further considered mostly relative voteshares $r_{iyp}$, which are absolute vote shares minus the national average for a given party every electoral year
\begin{eqnarray}
r_{iyp} = v_{iyp} - a_{yp}~, \label{relative_voteshare} \\
a_{yp} = \sum_i v_{iyp} / I ~,
\end{eqnarray}
 where I stands for the number of counties.

The relative voteshares show how much above (positive) or below (negative) the national average are the results in given county.\\
Our main focus is on the persistence of voting patterns. We define a \emph{stronghold} to be a county which relative voteshare remains systematically positive (or negative). 
%We can invert this definition, to define and calculate stronghold threshold value $t_{ip}$ for each county, that is the value of $t$ required for given county to be considered a stronghold. 
%It is equal to the lowest relative voteshare value across all time points (Figure \ref{example}), and is often negative. 
% We consider threshold $t=0$ to be ``default'' threshold when it is not mentioned and counties fulfilling the definition for this threshold a real stronghold for the party.\\
% Using the stronghold definition, we investigate percolation behavior, by looking at the clusters of strongholds. 
% We use physical adjacency (sharing any length of border, including single point) of counties to decide whether strongholds form a contiguous cluster or not.\\

\subsection{Model definition and analytical description}

In the SIRM model $N$ agents live in a spatial system divided in non-overlapping 
cells. The $N$ agents are distributed among the different cells according to 
their residence cell. The number of residents in a particular cell $i$ will be 
called  $N_i$. While many of these individuals may work at $i$, some others will 
work at different cells. This defines the fluxes $N_{ij}$ of residents of $i$ 
recurrently moving to $j$ for work. By consistency,  $N_{i}=\sum_j N_{ij}$. The 
working population at cell $i$ is $N'_i=\sum_j N_{ji}$ and the total population 
in the system (country) is $N = \sum_{ij}\, N_{ij}$. In this work, 
the spatial units correspond to the US counties and the population levels, 
$N_i$, and commuting flows, $N_{ij}$, are directly obtained from the 2000 census.

We describe agents' opinion by a  binary variable with possible values $+1$ or 
$-1$. The main variables are the number of individuals $V_{ij}$ holding opinion 
$+1$, living in county $i$ and working at $j$. Correspondingly, 
$V_i=\sum_lV_{il}$ stands for the number of voters living in $i$ holding opinion 
$+1$ and $V'_j=\sum_lV_{lj}$ for the number of voters working at $j$ holding 
opinion $+1$. We assume that each individual interacts with people living in her 
own location (family, friends, neighbors) with a probability $\alpha$, while 
with probability $1-\alpha$ she does so with individuals of her work place. Once 
an individual interacts with others, its opinion is updated following a noisy 
voter model 
\cite{Fowler2005,holley,vazquez_eguiluz,voter_redner,redner_kinetic}: an 
interaction partner is chosen and the original agent copies her opinion 
imperfectly (with a certain probability of making mistakes). A more detailed 
description of the SIRM model can be found in \cite{FernandezPRL}.

\section{Results}
\begin{figure}[]
   \epsfig{file=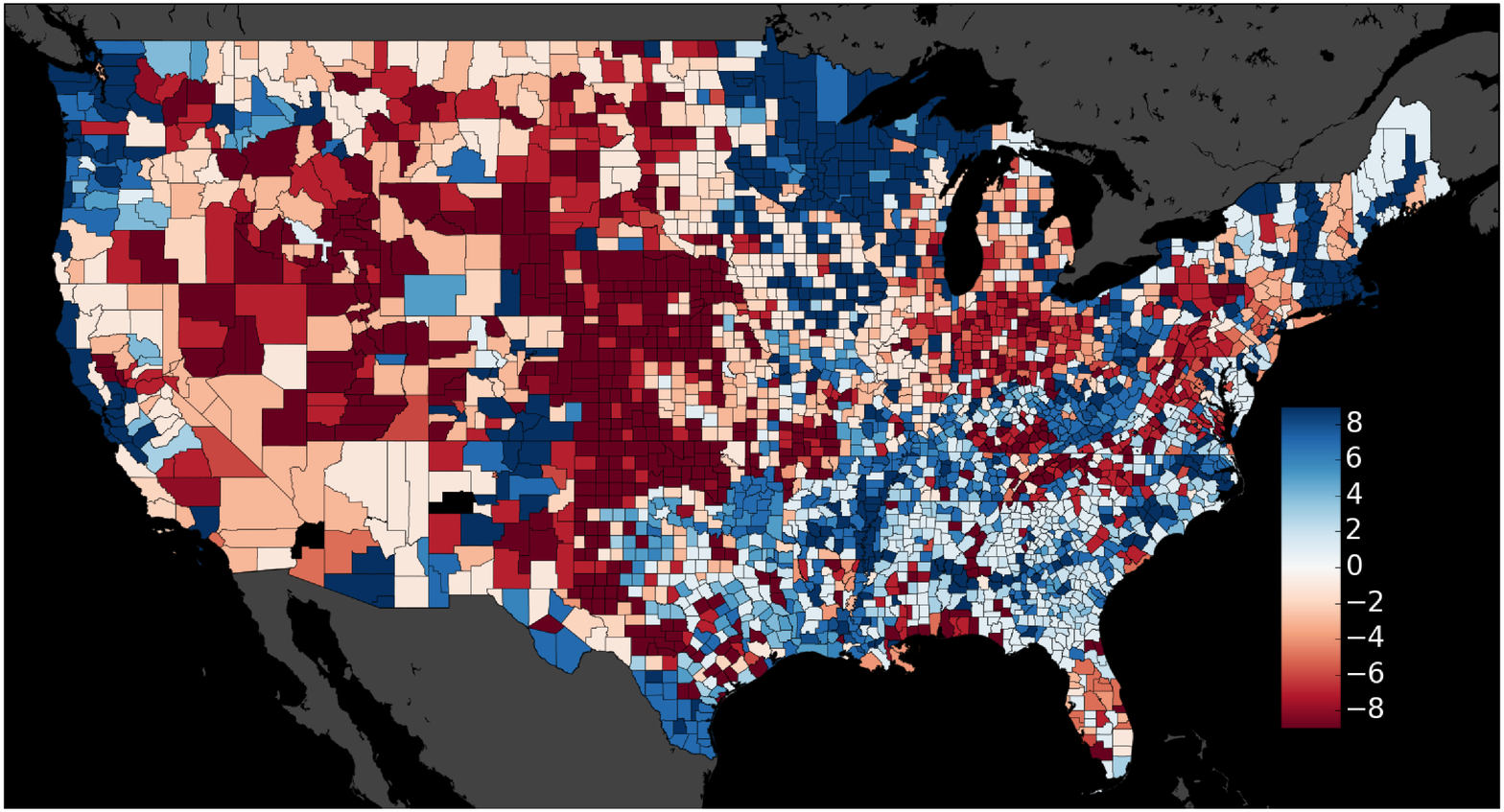, width=0.95\columnwidth}
   \epsfig{file=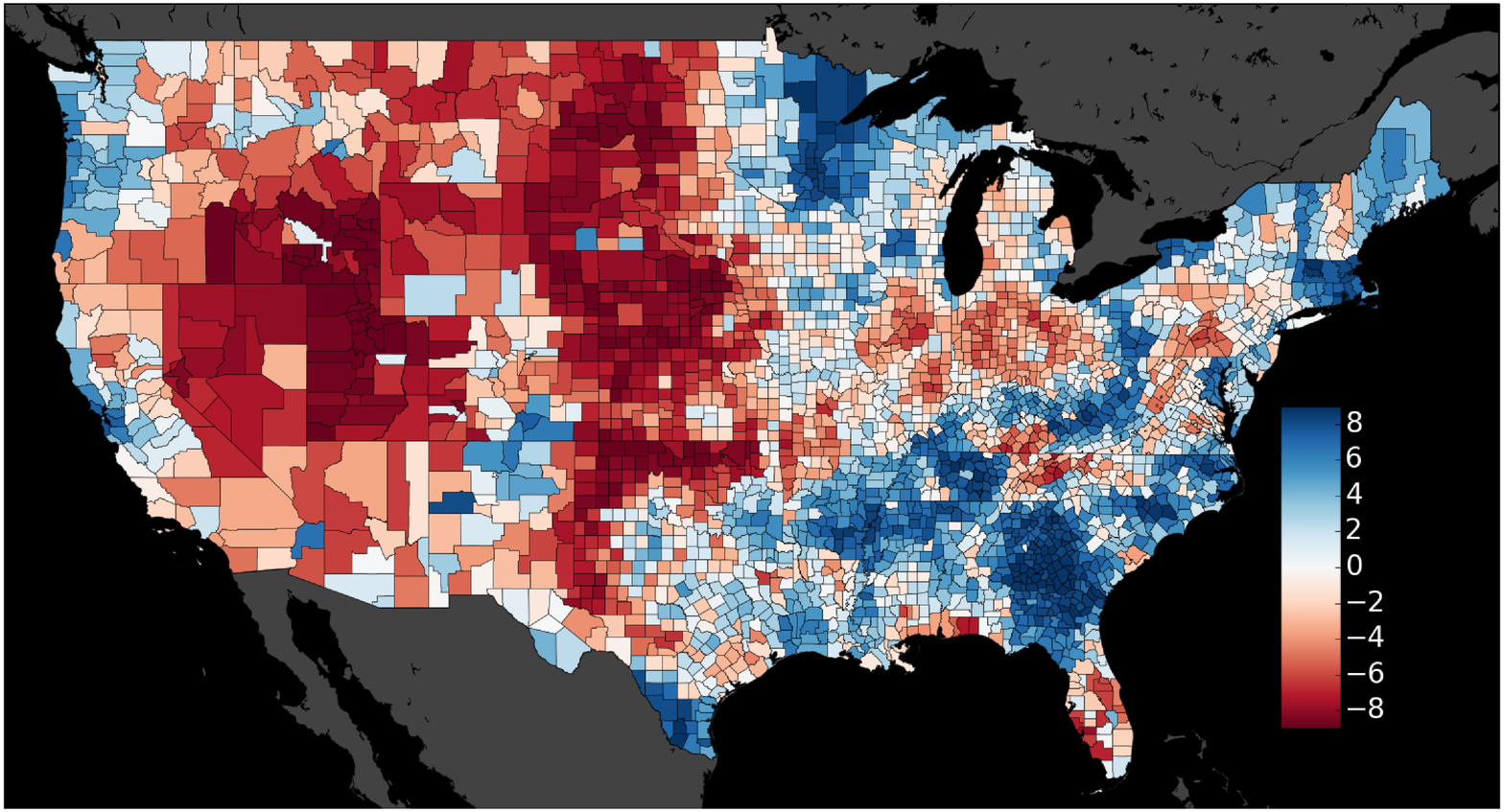, width=0.95\columnwidth}
   \caption{Spatial distribution of the strongholds (red shows republicans while blue shows democrats).  Codified by color the stronghold time measured in elections units 
   (republicans stronghold times are given by the absolute value of the color bar). 
    No county is a stronghold for both parties. Upper panel: US Presidential Elections data during the period 1980-2012. Bottom panel: model simulations for 9 elections using 
    the electoral results from 1980 (at county level) as initial condition.}
  \label{strongholds}
\end{figure}

Pursuing the topic of persistence and changes in opinion, as expressed by voting results, we have investigated the \emph{strongholds} 
of both parties in USA. Figure \ref{strongholds} shows the spatial arrangement and the duration of the strongholds (measured in elections) 
for data of the US Presidential Elections during the period 1980-2012 and simulations using the electoral results of 1980 as initial condition. 
It can be seen at a glance, that the strongholds are not randomly distributed across the country, but clustered. This indicates that some form of correlation 
is present between the voting patterns. The republican strongholds seem to be concentrated mostly in the central-west, while democrat 
strongholds are dispersed mostly through the eastern parts, including urbanized areas. This is in agreement with the population distributions, 
republican strongholds being mostly lower populated counties, while democrats strongholds include some significant cities.

\begin{figure}[]
   \epsfig{file=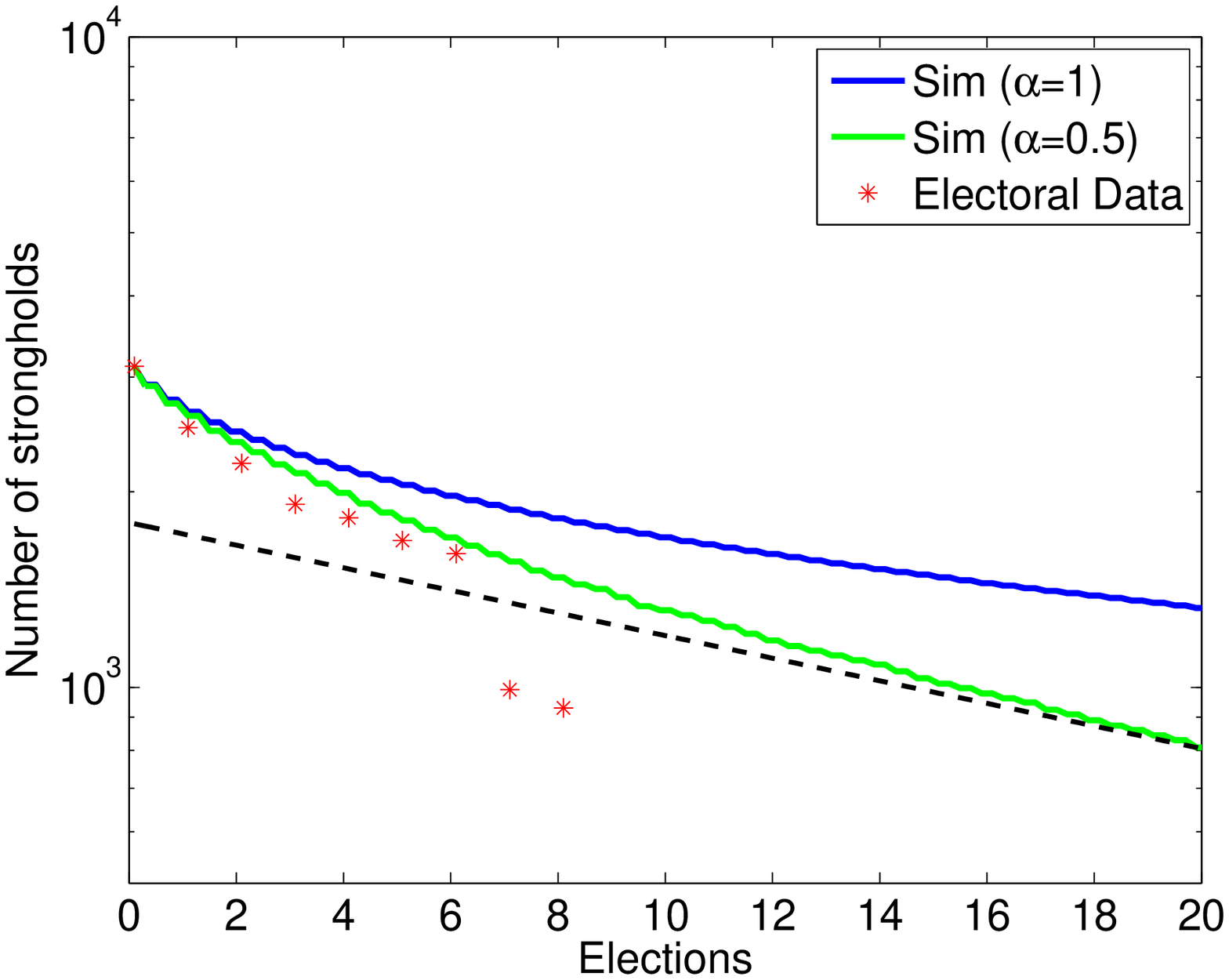, width=0.95\columnwidth}
   \caption{Temporal evolution of the number of strongholds for electoral data and simulation results. Two different simulation results are presented, for $\alpha=0.5$ 
   the individuals have the same probability of interacting with individuals from other county as with individuals of his own county, while for $\alpha=1$ interactions only 
   within the same county would happen. 
   Dashed line fits $f(t)=\frac{N_c}{2}\exp(-t/b)$ with $b=25$ elections.}
  \label{strongholds_decay}
\end{figure}

We quantify in Figure \ref{strongholds_decay} the temporal evolution of the strongholds of the election data for the period 1980-2012 as well as for the strongholds forecasted 
by simulations after 9 elections. 
The influence of the commuting network and their interactions was revealed by contrasting the simulations results with and without network interaction. 
The evolution of the number of strongholds observed in the data at early stages is well described by the model with commuting interactions. 
For longer times, the model predicts that the number of strongholds will decay with time following an exponential law as shown in Figure \ref{strongholds_decay}. 
In the absence of commuting interaction, the number of strongholds decreases at a slower rate. Furthermore, the model overestimate the number of strongholds in this case
since no other mechanisms than internal fluctuations act driving the county relative voteshare $r_{iyp}$ towards the average. Counties set as strongholds at the beginning of 
the simulation remains strongholds for a longer time.

\begin{figure}[]
   \epsfig{file=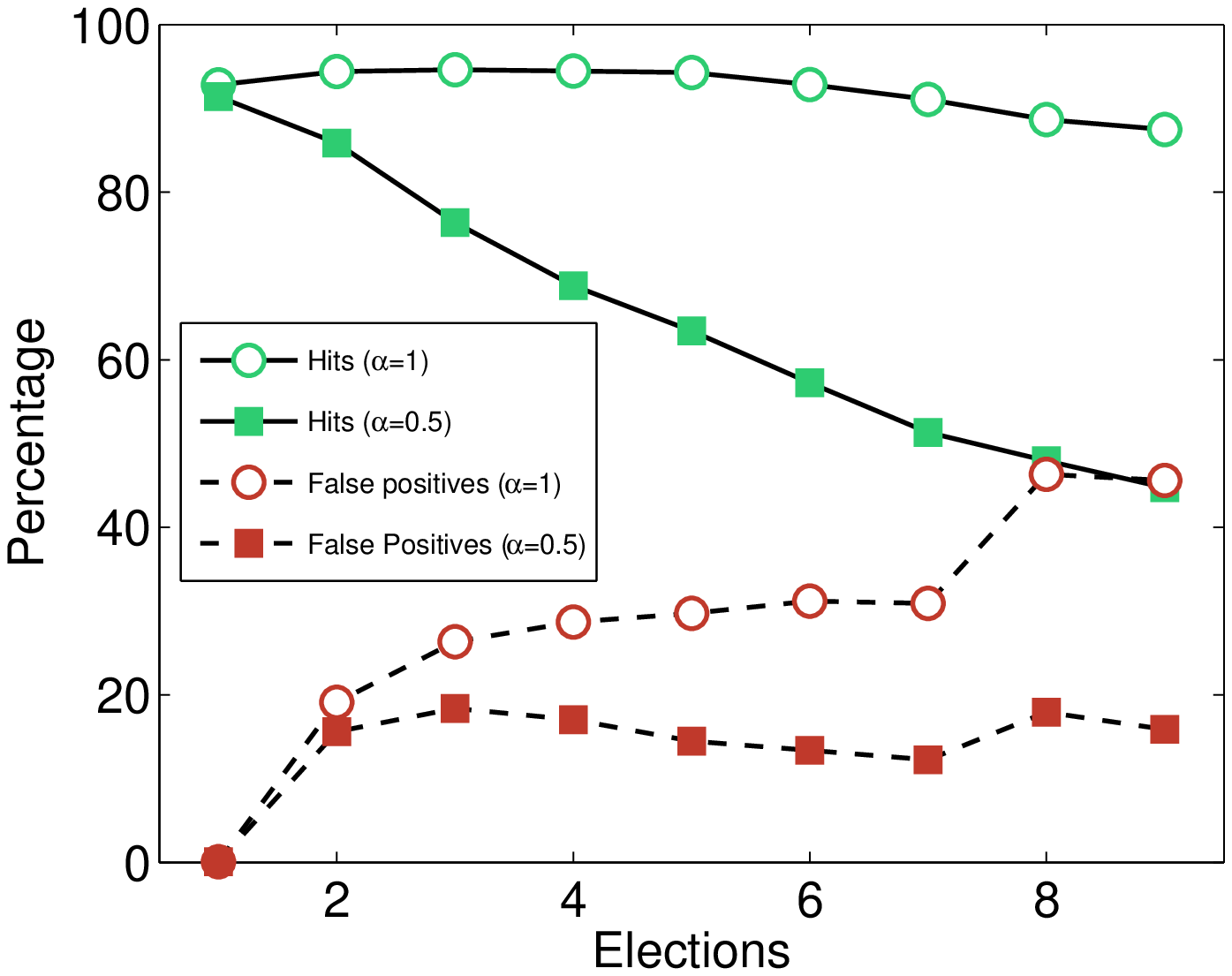, width=0.95\columnwidth}
   \caption{Variation of the percentage of strongholds accurately predicted by the model (hits) for several elections. The number of false positives, i.e., counties identified 
   as strongholds in the simulations without stronghold correspondence in the data is also shown.}
  \label{percenthits}
\end{figure}

We further test the accuracy of our model by computing the percentage of strongholds accurately predicted. As Figure \ref{percenthits} shows, the model with no commuting interactions 
reproduces a higher percentage of strongholds, however, it also gives higher and increasing number of false positives. On the contrary, the model with commuting interactions maintain 
a flat rate of false positive below 20\%. The accuracy of the model in this case remain higher than 50\% after $7$ elections.

\begin{figure}[]
   \epsfig{file=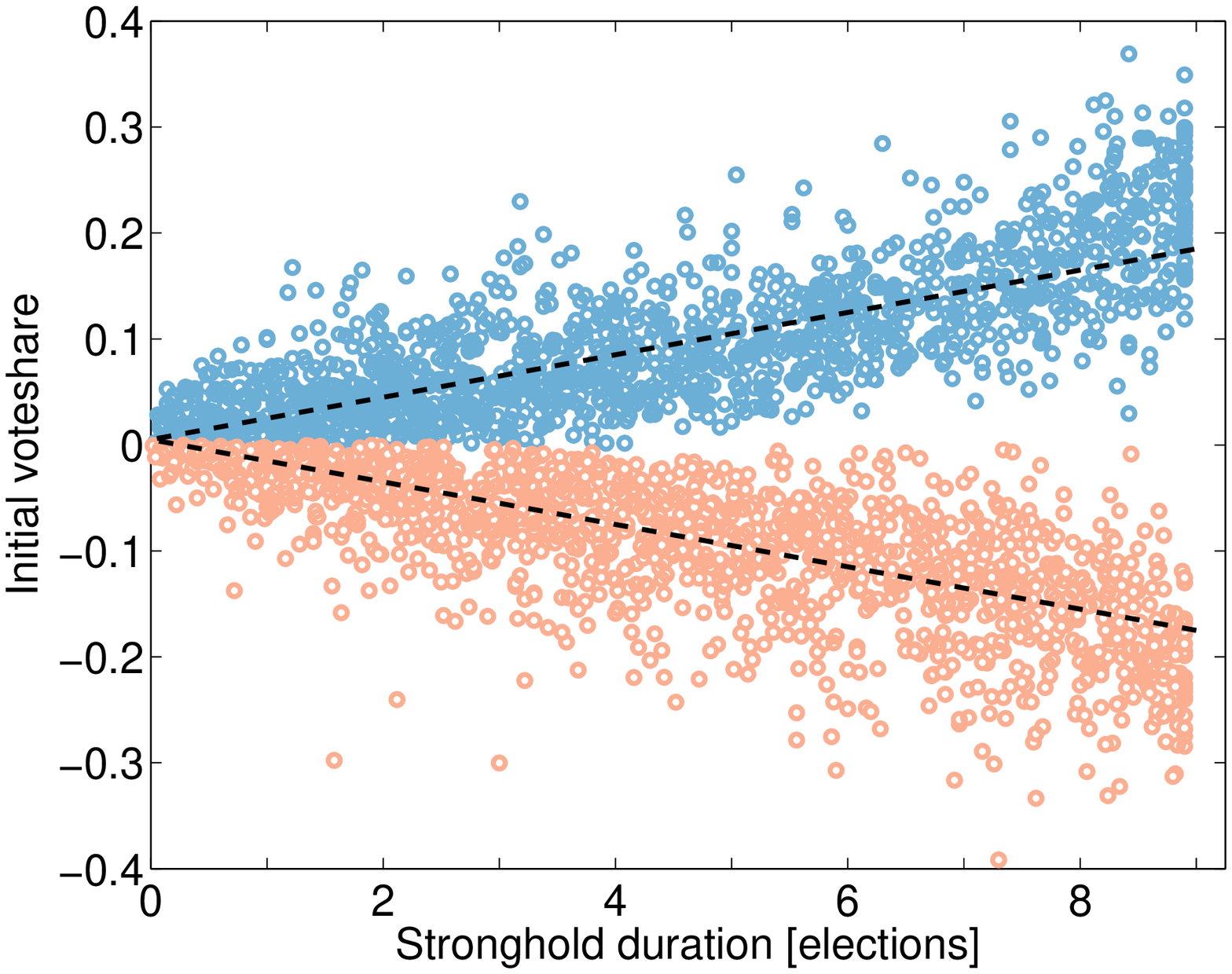, width=0.95\columnwidth}
   \caption{Scatter plot of the initial voteshare versus the stronghold duration for model results. Red symbols shows republicans while blue symbols shows democrats.  
   Dashed line correspond to the least square fitting revealing a symmetric linear behavior $y=mx+n$ with $m=\pm 0.02$ and $n=\pm 0.005$.}
  \label{strongholds_CIdependence}
\end{figure}

Another question is what determines how long a county would be a stronghold for? We try to answer this question for the model by looking into the dependence of the strongholds duration 
with respect to the initial voteshare of the counties. As Figure \ref{strongholds_CIdependence} shows, there is a linear dependence for the duration of being a stronghold with the 
distance to the mean voteshare. Counties with initial larger deviation from the mean voteshare tend to be strongholds longer time. The fitting reveals that every tenth of 
relative voteshare corresponds on average to a stronghold duration of $5$ elections. \\

\section{Discussion}
We have studied the persistence on the electoral system using the recently introduced 
Social Influence and Recurrent Mobility (SIRM) model for opinion dynamics which includes social influence
with random fluctuations, mobility and population heterogeneities across the U.S. 
The model accurately predicts generic features of the background fluctuations of evolution of vote-share
fluctuations at different geographical scales (from the county to the state level), but it does not
aim at reproducing the evolution of the average vote-share. 

We have contrasted the evolution of the number of strongholds of the election data for the period 1980-2012 with the strongholds forecasted by simulations 
finding a good agreement between them. The evolution of the number of strongholds observed in the electoral data at early stages is well described by the model with commuting interactions. 
However, the number of strongholds observed in the data changes abruptly after the presidential election of 2008. Our model is not able/does not intend to reproduce this behavior since 
it involves external driving forces not included in the model. 

Our results also shows a good agreement between data and simulations for the location and duration of the strongholds. 
Strongholds are not randomly distributed across the country, but clustered. This indicates that some form of correlation 
is present between the voting patterns. The model reproduces nicely the spatial concentration of the both types of strongholds. 
Republican strongholds are mostly concentrated in the central-west, while democrat strongholds are dispersed mostly through the eastern parts, including urbanized areas. 
 
As for the duration of the strongholds, we have found a linear dependence for the duration of being a stronghold with the 
distance to the mean voteshare. Counties with initial larger deviation from the mean voteshare tend to be strongholds for a longer time, on average, 5 elections for each tenth of relative voteshare. 
When commuting interactions are taken into consideration, our model exhibits an accuracy higher than 50\% for up to $7$ elections. 
The lack of this interactions causes an overestimation of the number of strongholds with an associated increase of the number of false positive 
decreasing the prediction accuracy of the model. 

Our contribution sets the ground to include other important aspects of voting behavior and 
demand further investigation of the role played by heterogeneities in the micro-macro connection. 
Further elements will have to be included in order to produce predictions mimicking more accurately real electoral results. 
Some examples are the effects of social and communication media or the erosion of the governing party. 
The use of alternative communication channels is expected to affect voting behavior.

\subsubsection{Acknowledgements.}
The authors acknowledge support from project MODASS (FIS2011-24785). 
TP acknowledges support from the program Juan de la Cierva of the Spanish Ministry of Economy and Competitiveness.

\end{document}